\documentclass[11pt,a4paper]{article}
\usepackage{jheppub}

\usepackage{physics}
\usepackage{tensor}
\usepackage{amsmath}

\numberwithin{equation}{section} 
\usepackage{amssymb}
\usepackage[colorlinks=true]{hyperref}
\usepackage{xcolor}
\usepackage{cancel}
\usepackage{ytableau}

\usepackage{titlesec}
\makeatletter
\g@addto@macro\bfseries{\boldmath}
\makeatother

\ytableausetup{boxsize=0.3em,centertableaux}

\usepackage{framed}
\usepackage[most]{tcolorbox}
\usepackage{xcolor}
\colorlet{shadecolor}{gray!10}
\usepackage{marginnote} 

\newcommand{\re}[1] {(\ref{#1})}

\newcommand{\nn}{\nonumber}
\newcommand{\al}{\alpha} 
\newcommand{\be}{\beta}

\newcommand{\beq}{\begin{equation}}
\newcommand{\eeq}{\end{equation}}

\preprint{UUITP-30/26} 

\title{Gauging background isometries of the ILST null string
}

\author[a]{Ulf Lindstr\"{o}m}

\affiliation[a]{Department of Physics and Astronomy, Uppsala University,
Box 516, SE-75120 Uppsala, Sweden
and Centre for Geometry and Physics, Uppsala University,
Box 480, SE-75106 Uppsala, Sweden
}

\emailAdd{ ulf.lindstrom@physics.uu.se}

\abstract{We present a general gauging of a subalgebra, $\mathfrak h$, of the isometries $\mathfrak g$ of the background geometry for the ILST string \cite{Isberg:1993av}. We discuss which of the isometries in $\mathfrak g$ survive the gauging as symmetries. The gauged symmetry does not in general involve transformations of the vector density $V^\al$. There is thus a corresponding contraction of the algebra of constraints as compared to the flat background and it no longer represent the full Carroll-Weyl algebra. This algebra of constraints is derived and discussed, partly in an appendix. The gauging of isometries sets the stage for a reduction, or quotient construction, leading to a new null string on the space of orbits of the gauge group. We illustrate this by deriving the ILST string in $AdS$ background as a quotient of a model in $SO(d,2)$ background by a $SO(d,1)$ isometry, in group space.}

\begin{document}
\pagestyle{myplain}
\maketitle

\section{Introduction}
 Strings with a null world sheet were introduced a long time ago, from two slightly different points of view \cite{Schild:1976vq,Karlhede:1986wb}. They were given a more geometric guise in \cite{Isberg:1993av}, recently named the ILST string, and have attracted some interest lately. 
The action for this string is
 \beq 
\label{ILST}   
S=\int d^2\xi V^\al {\partial}_{\al} X^m\eta_{mn}V^\be{\partial}_{\be}  X^n~,
\eeq
 where the world sheet coordinates are $\xi^\al=(\tau,\sigma)$, $V^\al(\xi)$ are  worldsheet vector densities, $X^m,~m=0,...,d-1$ and $\eta$ are the ambient spacetime coordinates and metric.  

 Early on it was observed that the model with metric $\eta$ is invariant under the conformal group of spacetime transformations with compensating transformations of the $V$s, see eg. \cite{Isberg:1993av}. The dilatation part reads
 \beq\label{TR1}
 \delta X^m=\lambda X^m~,~~\delta V^\al=-\lambda V^\al~,
 \eeq
 with $\lambda$ a constant parameter.
A  number of studies are concerned with gauging this dilatation symmetry to allow for a $\lambda(\xi)$, beginning with \cite{Gustafsson:1994kr}. Recent examples are  \cite{Sheikh-Jabbari:2026tpf,Sheikh-Jabbari:2026vqh,Duary:2026lmk,Duary:2026zar,Lindstrom:2026zno,Lindstrom:2026tua,Lindstrom:2026jbi}.
The gauged version of \re{ILST} reads
 \beq\label{GILST}
 \int d^2\xi \big(V^\al  {\cal D}_{\al}  X^{m} \eta_{mn}V^\be{\cal D}_{\be} X^{n}\big)
 \eeq
 with covariant derivative ${\cal D}_{\al}  X^{m}:=\partial_\al X^m-W_\al X^m$. This action is invariant under the local version of \re{TR1} together with the gauge field transforming as $\delta W_\al=\partial_\al \lambda$. 

 In view of what follows, note that this is not a gauging of an isometry of $\eta$ but rather of a homothety; the Minkowski metric scales with a constant $\lambda$ under $\delta X^m=\lambda X^m$.

A Hamiltonian analysis of the local scaling symmetry of \re{GILST} leads to a semi direct product of 
the Virasoro algebra with a su(1,1) Kac-Moody algebra \cite{Gustafsson:1994kr} or a Carroll-Weyl algebra \cite{Sheikh-Jabbari:2026vqh}, which accounts for the present interest.

\section{Gauging background isometries}
\label{iso}

A natural question to ask is how the model and its symmetries change in a curved background. This has been addressed, mainly for $AdS$ background, in \cite{Uvarov:2017lyu,Duary:2026zar,Duary:2026izp,Lindstrom:2026tua,Lindstrom:2026jbi}.

There are no homothety symmetries in $AdS$ and $dS$, so to mimic \re{GILST} we either have to find a modified background or gauge an isometry. Choosing the latter route we consider 
\beq  
\label{BILST}
S=\int d^2\xi V^\al {\partial}_{\al} X^m G_{mn}(X)V^\be{\partial}_{\be}  X^n~.
\eeq 
It is clear from the close semblance to a non-linear sigma model that this action is invariant under isometries of the (arbitrary) metric $G_{mn}$. This also makes it clear how to gauge the isometry \cite{Hull:1985pq}.

Assume that $G$ has a Killing vector\footnote{An interesting generalization, suggested by B. Sundborg, is to gauge conformal isometries.} $k^m$, then
\beq\label{XTf}
\delta X^m=k^m(X)=kX^m=k^n\partial_n X^m~,~~\makebox{and the Lie derivative}~~~~{\cal L}_{\lambda k} ~G(X)=0~.
\eeq
More generally, assume that $G$ has $n$ Killing vectors $k^m_A, ~A=1,...,n$ that form an algebra
\beq
[k_A,k_B]=f_{AB}^{~~C}k_C~.
\eeq
Then the covariant derivative is
\beq
\nabla_\al X^m:=\partial_\al X^m-W_\al^Ak^m_A
\eeq
where the gauge field $W_\al^A$ transforms infinitesimally as
\beq
\delta W_\al^A=\partial_\al\lambda^A+f_{CB}^{~~A}W^C_\al\lambda ^B
\eeq
which implies that 
\beq\label{NaTf}
\delta \nabla_\al X^m=\lambda^Ak^m_A,{}_n\nabla_\al X^n~.
\eeq
Replacing partial by covariant derivatives in \re{BILST} we have
\beq\label{GBILST} 
S_G=\int d^2\xi V^\al {\nabla}_{\al} X^m G_{mn}(X)V^\be{\nabla}_{\be}  X^n~.
\eeq
Using \re{NaTf} and  \re{XTf} we verify that $S_G$ is invariant with $V^\al$ inert.\footnote{In the special case of the $AdS$ metric in Poincar\'e coordinates and a single scale transformations of $X$, the metric is homogeneous of degree $-2$ under the scaling and invariance of (\ref{BILST}) follows due to Euler's homogeneous function theorem. The $V$ field is again inert. This seems different from the gauging procedure, but it is in fact just that construction for the special case when the Killing vector is $k^m=X^m$. In general coordinates the schedule becomes exactly the gauging described above. } 

Not all global isometries of $(\mathcal M,G)$ survive as symmetries of the gauged action. Let $\xi$ be an isometry, $\delta X^m=\epsilon\,\xi^m(X)$, and let the gauged algebra $\mathfrak h\subset\mathfrak{iso}$ be spanned by the $k_A$, as above. Invariance of $,\nabla X^mG_{mn}\nabla X^n$ requires $\delta(\nabla X^m)=\epsilon\,\partial_n\xi^m\,\nabla X^n$, which implies
\beq
\delta W^A\,k_A^m=\epsilon W^A\,[k_A,\xi]^m .
\eeq
This can be solved for $\delta W$ only if $[k_A,\xi]\in\mathrm{span}\{k_C\}$ for all $A$. Hence the global symmetry of the gauged model is the normalizer $N(\mathfrak h)$ of $\mathfrak h$ in $\mathfrak{iso}$. Those elements of $N(\mathfrak h)$ lying in $\mathfrak h$ are gauge transformations, so the physical global symmetry is $N(\mathfrak h)/\mathfrak h$. Isometries outside $N(\mathfrak h)$ are symmetries of the ungauged model but are broken by the gauging. They map the gauged model to a different one, built on a conjugate $\mathfrak h$. Geometrically, $N(\mathfrak h)$ generates the isometries that map $H$-orbits to $H$-orbits, so only these will descend to the quotient ${\mathcal M}/H$ in the quotient construction of sections \ref{Quote} and \ref{AdSquote}.

\section{Constraint algebra}

A Hamiltonian analysis of \re{GBILST} will result in a set of constraints and their algebra. For the case of a single Killing vector we find $\mathrm{BMS}_3$ extended by an abelian weight-one $u(1)$ current, with $[M,J]=0$, $J$ being  the Fourier modes of
${\cal C}_{PX}=P\cdot k(X)$. A particular case is gauging the Killing vector $
k=z\partial_z+x^\mu\partial_{\mu}$ with the $AdS$ metric in Poincar\'e coordinates. Cf. the gauging of a homothety vector in $AdS$ in \cite{Duary:2026zar}. 

A noteworthy fact is that the algebra of constraints \re{Contracted} is formally the same as that for the $eAdS$ string described in  \cite{Lindstrom:2026jbi}, in spite of the fact that the $eAdS$ string represents a very different gauging and that there is no map between them.

When gauging a (sub)algebra of isometries $\mathfrak h$, the first-class constraints close to a $\mathrm{BMS}_3$
–Kac–Moody algebra, i.e. $\mathrm{BMS}_3$ extended by the loop algebra of $\mathfrak h$ acting by weight-one currents that commute with the supertranslations.

\section{Quotients}
\label{Quote}

The quotient construction entails gauging a symmetry and then finding a special gauge connection that extremises the action. Using this gauge connection in the action yields a model on the space of orbits of the gauge group. In the form of the Marsden-Weinstein symplectic quotient and/or the Hyperk\"ahler reduction it plays an important role in differential geometry \cite{Hitchin:1986ea}.  Here we present the procedure as applied to \re{GBILST}.
The variational equation for $W_\al^A$ is   
\begin{align}\nn
&-k^m_A G_{mn}V\partial X^n+k^m_A G_{mn}k^n_BVW^B=0\\[2mm]
&\Rightarrow VW^A=\mathbb{H}^{AB}k_{Bn}V\partial X^n~,
\end{align}
where $\mathbb{H}_{AB}:=k^m_A G_{mn}k^n_B$ and $\mathbb{H}^{AB}\mathbb{H}_{BC}=\delta^A_C$. It follows that
\beq
S_G\to \int d^2\xi V^\al {\partial}_{\al} X^m \hat G_{mn}(X)V^\be{\partial}_{\be}  X^n~,
\eeq
where the quotient metric is
\beq
\hat G_{mn}:=G_{mn}-k_{Am}\mathbb{H}^{AB}k_{Bn}~.
\eeq

\section{The ILST string in AdS background as a quotient.}
\label{AdSquote}
 In this section we adapt a well known construction from sigma models to exemplify the quotient.
 
Lorentzian $AdS_{d+1}$ is the harmonic space ${SO(d,2)}/{SO(d,1)}$. As an illustration of the quotient construction of null strings in group spaces, we use this fact to derive the ILST string in an $AdS$ background\footnote{In \cite{Lindstrom:2026jbi} the same result is achieved via the embedding formalism}.

 In the group space construction of sigma models the basic element is a group element $g\in G$. We parametrize the group elements using coordinates $X^m$,
 and define the left $L$ and right $R$ frames by
 \beq
 g^{-1}\delta g=L_m\delta X^m~,~~~\delta gg^{-1}=R_m\delta X^m~.
 \eeq
 The frames are Lie algebra valued: $L_m=:L_m^A T_A~,~R_m=:R_m^A T_A$ with $T_A$ Lie algebra generators. To define a sigma model on group space we need the Lie algebra valued Maurer-Cartan form  $J_\al=g^{-1}\partial_\al g$. Then the relation to the previously discussed form of the ILST string action is
 \beq
 \int d^2\xi V {\partial} X^m G_{mn}(X)V{\partial} X^n={\textstyle \frac 1 2}Tr\int d^2\xi <J_VJ_V>
 \eeq
 which equals $\int d^2\xi VJ^AG_{AB}VJ^B$ 
 when $G_{AB}$ is the Killing metric 
 ${\textstyle \frac 1 2}Tr(T_AT_B)$. With $H=SO(d,1)$ a subgroup of $G=SO(d,2)$ the Lie algebra splits as $\mathfrak g=\mathfrak h\oplus\mathfrak m$. 
 We gauge $H$ on the right introducing $w$ by letting $J_V\to J_V+w$ with $w=VW\in \mathfrak h$.
 
Eliminating $w$  projects onto the coset directions:
\beq
S=\tfrac12\int\mathrm{Tr}\big((J_VJ_V)_{\mathfrak m}\big),
\eeq
This uses the orthogonality of $\mathfrak h$ and $\mathfrak m$. The result is ILST on $G/H=AdS_{d+1}$, the quotient metric $\hat G$ for this case is re-derived in the process.

The relevant part of the discussion in section \ref{iso} is that here $\mathfrak{iso}=\mathfrak g_L\oplus\mathfrak g_R$ and $\mathfrak h\subset\mathfrak g_R$. All of $\mathfrak g_L$ lies in $N(\mathfrak h)$ because left and right translations commute, so $G_L$ survives. In $\mathfrak g_R$ only $N_{\mathfrak g_R}(\mathfrak h)$ survives, which is $\mathfrak h$ itself when $\mathfrak g$ is simple. So the physical symmetry is $G_L$, the isometry of $G/H$.
\vspace{20mm}

\noindent
{\bf Acknowledgement}\\
\noindent
It is a pleasure to acknowledge numerous useful discussions with Bo Sundborg on the topics of this paper.
\bigskip

\noindent
{\bf Use of artificial intelligence}\\
\noindent
Claude, Anthropic, was used for literature search in the preparation of this work.

\vspace{20mm}

\Large
\centerline {\bf Appendix}
\normalsize
\bigskip
\appendix

\section{ AdS  and some of its coordinates}
\label{AdsGeom}
This appendix is essentially reproduced from our paper \cite{Lindstrom:2026jbi}.

Anti de Sitter space in $d+1$ dimensions can be realized as a homogeneous space, the coset space 
\beq
 AdS_{d+1}=\frac {SO(d,2)}{SO(d,1)}~.
\eeq{}
We will make use of an explicit construction adapted from Dirac's conformal space. So consider $\mathbb{R}^[d.2]$  with coordinates $X^M$, $M=0,1,...,d,d+1$ and metric signature $(-,+++...+,-)$. Imposing the embedding constraint 
\beq\label{emb}
X^2=\sum_{m=1}^{d}(X^m)^2-(X^0)^2-(X^{d+1})^2=-R^2
\eeq
reduces the dimension to $d+1$.

Solving \re{emb} can be done using different coordinates. Perhaps the most direct solution uses ``Beltrami coordinates''. We give them below and their relation to the  more well known Poincar\'e coordinates.

We  may reduce from $d+2$ to $d+1$ by solving  the embedding condition for $X^{d+1}$:
\beq\label{weier}
X^{d+1} = \pm \sqrt{X^2+R^2}~, 
\eeq
where $X^2:=X^mX^n\eta_{mn}$, $m,n=0,1,...d$. The induced $d+1$ dimensional line element then reads
\beq\label{Winduced}
ds_{|}^{2}=dX^m\Big(\eta_{mn} -\frac{X_{m}X_{n}}{X^{2}+R^{2}}\Big)dX^n~ =: dX^m\hat\eta_{mn}dX^n~,
\eeq
which is recognizable as the metric on the space of orbits for the isometry, the bulk isometry that scales the vector \(X^{d+1} + X^0 \). This is an example of a reduction (quotient) under an isometry. 

The relation to the Poincar\'e coordinates is 
\begin{align}
&X^{d+1} + X^0 = \frac{R^2}{z}\\\nn
&X^{d+1} - X^0 = -z - \frac{\eta_{\mu\nu}x^\mu x^\nu}{z}\\
&X^\mu = \frac{R x^\mu}{z} ~,
\end{align}

where (\(\mu = 1, \dots, d\)).

\section{The constraint algebra for the ILST with one isometry gauged}

Consider the action \re{GBILST} with one isometry of the metric $G$ gauged. Denote its Killing vector $k^m(X)$. The constraints that follow from the gauged model are then
\beq\label{redconsts}
{\cal C}_1=G^{mn}P_mP_n~,~~ {\cal C}_2=P\cdot\partial_\sigma X:=P\cdot X'~,~\makebox{and} ~~{\cal C}_{PX}=P\cdot k(X)~.
\eeq
where
\beq
 P_m=2G_{mn}V^\tau V\cdot {\nabla} X^n~,
 \eeq
i.e. they look identical to those in an $\eta_{mn}$ background, except that $G^{mn}P_mP_n$ now contains an $X^m$ dependent inverse metric. 
Due to the presence of a curved background
we expect the Carroll-Weyl symmetry they generate to become modified. 
Since the string is moving in a curved background one possibility is to turn to covariant phase space notation. However, we  use standard Poisson brackets and carefully keep track of all additional functions when evaluating Poisson brackets.
The crucial bracket when comparing to the $\eta_{mn}$ background is 
$[\mathcal C_{PX}(\sigma),\mathcal C_1(\sigma')]=0$.
The curved background geometry leaves all the other brackets intact, but the fact that the constraint $G^{mn}P_mP_n$ no longer scales affects the algebra of constraints which now reads: 

\begin{align}\nn
\label{Contracted}
&[\mathcal C_1(\sigma),\mathcal C_1(\sigma')]=0\\\nn
&[\mathcal C_2(\sigma),\mathcal C_2(\sigma')]=(\mathcal C_2(\sigma)+\mathcal C_2(\sigma'))\delta'(\sigma-\sigma')\\\nn
&[\mathcal C_{px}(\sigma),\mathcal C_2(\sigma')]=\mathcal C_{px}(\sigma')\delta'(\sigma-\sigma')\\\nn
&[\mathcal C_{PX}(\sigma),\mathcal C_{PX}(\sigma')]=0\\\nn
&[\mathcal C_1(\sigma),\mathcal C_2(\sigma')]=(\mathcal C_1(\sigma)+\mathcal C_1(\sigma'))\delta'(\sigma-\sigma')\\
&[\mathcal C_{PX}(\sigma),\mathcal C_1(\sigma')]=0~.
\end{align}

 Down to the last line these are the relations of the Carroll-Weyl symmetry, but the inclusion of that last relation restricts to a contraction/deformation of the Carroll–Weyl algebra in which the "Weyl weight" coupling between scalings and the Hamiltonian has been switched off, a ``Static'' variant.
 At the level of Fourier modes the statement is that the commutator $[M_m,J_n]$ vanishes, where $M_m$ corresponds to supertranslations and $S
J_n$ to the Weyl/rescaling current. See, e.g., \cite{Lindstrom:2026zno}. 
 \bigskip

\bibliographystyle{JHEP}
\bibliography{references}

@article{Lindstrom:2026tua,
    author = {Lindstr{\"o}m, Ulf and Sundborg, Bo},
    title = "{(Anti-)De Sitter null strings and Carroll-Weyl symmetry}",
    eprint = "2609.07765",
    archivePrefix = "arXiv",
    primaryClass = "hep-th",
    reportNumber = "UUITP-21/26",
    month = "9",
    year = "2026"
}

@article{Lindstrom:2026jbi,
    author = {Lindstr{\"o}m, Ulf and Sundborg, Bo},
    title = "{eAdS- embedded AdS, tensionless strings and contracted Carroll-Weyl symmetry}",
    eprint = "2611.10907",
    archivePrefix = "arXiv",
    primaryClass = "hep-th",
    reportNumber = "Uppsala Theoretical Physics report UUITP-27/26",
    month = "11",
    year = "2026"
}

@article{Lindstrom:2026zno,
    author = {Lindstr{\"o}m, Ulf},
    title = "{The conformal null string in $d+2$ and $d$ dimensions}",
    eprint = "2606.22498",
    archivePrefix = "arXiv",
    primaryClass = "hep-th",
    reportNumber = "Uppsala University, Theoretical Physics UUITP-15/26",
    month = "6",
    year = "2026"
}

@article{Gustafsson:1994kr,
    author = "Gustafsson, H. and Lindstr{\"o}m, U. and Saltsidis, P. and Sundborg, B. and von Unge, R.",
    title = "{Hamiltonian BRST quantization of the conformal string}",
    eprint = "hep-th/9410143",
    archivePrefix = "arXiv",
    reportNumber = "USITP-94-08",
    doi = "10.1016/0550-3213(95)00051-S",
    journal = "Nucl. Phys. B",
    volume = "440",
    pages = "495--520",
    year = "1995"
}

@article{Isberg:1993av,
    author = "Isberg, J. and Lindstr{\"o}m, U. and Sundborg, B. and Theodoridis, G.",
    title = "{Classical and quantized tensionless strings}",
    eprint = "hep-th/9307108",
    archivePrefix = "arXiv",
    reportNumber = "USITP-93-12",
    doi = "10.1016/0550-3213(94)90056-6",
    journal = "Nucl. Phys. B",
    volume = "411",
    pages = "122--156",
    year = "1994"
}

@article{Karlhede:1986wb,
    author = "Karlhede, A. and Lindstr{\"o}m, U.",
    title = "{The Classical Bosonic String in the Zero Tension Limit}",
    doi = "10.1088/0264-9381/3/4/002",
    journal = "Class. Quant. Grav.",
    volume = "3",
    pages = "L73--L75",
    year = "1986"
}

@article{Schild:1976vq,
    author = "Schild, Alfred",
    title = "{Classical Null Strings}",
    reportNumber = "PRINT-76-0491 (TEXAS), ANL-HEP-PR-77-23",
    doi = "10.1103/PhysRevD.16.1722",
    journal = "Phys. Rev. D",
    volume = "16",
    pages = "1722",
    year = "1977"
}

@article{Duary:2026lmk,
    author = "Duary, Sarthak and Maji, Sourav",
    title = "{BRST quantization of Carroll-Weyl gauged null strings}",
    eprint = "2608.02731",
    archivePrefix = "arXiv",
    primaryClass = "hep-th",
    month = "8",
    year = "2026"
}

@article{Duary:2026izp,
    author = "Duary, Sarthak and Maji, Sourav",
    title = "{Null string holography in AdS from local BRST descent}",
    eprint = "2610.10540",
    archivePrefix = "arXiv",
    primaryClass = "hep-th",
    month = "10",
    year = "2026"
}

@article{Duary:2026zar,
    author = "Duary, Sarthak and Maji, Sourav",
    title = "{Carroll-Weyl gauged null strings in AdS}",
    eprint = "2610.10717",
    archivePrefix = "arXiv",
    primaryClass = "hep-th",
    month = "10",
    year = "2026"
}

@article{Sheikh-Jabbari:2026vqh,
    author = "Sheikh-Jabbari, M. M. and Yavartanoo, H.",
    title = "{Null Strings Gauged and Reloaded, I: Null Strings Have Carroll-Weyl Gauge Symmetry}",
    eprint = "2605.25817",
    archivePrefix = "arXiv",
    primaryClass = "hep-th",
    month = "5",
    year = "2026"
}

@article{Sheikh-Jabbari:2026tpf,
    author = "Sheikh-Jabbari, M. M. and Yavartanoo, H.",
    title = "{Null Strings Gauged and Reloaded, II: Consistent Classical Treatment of the Null Strings}",
    eprint = "2605.26822",
    archivePrefix = "arXiv",
    primaryClass = "hep-th",
    month = "5",
    year = "2026"
}

@article{Hitchin:1986ea,
    author = "Hitchin, Nigel J. and Karlhede, A. and Lindstrom, U. and Rocek, M.",
    title = "{Hyperkahler Metrics and Supersymmetry}",
    reportNumber = "STOCKHOLM-ITP-86-8",
    doi = "10.1007/BF01214418",
    journal = "Commun. Math. Phys.",
    volume = "108",
    pages = "535",
    year = "1987"
}

@article{Hull:1985pq,
    author = "Hull, C. M. and Karlhede, A. and Lindstr{\"o}m, U. and Ro{\v c}ek, M.",
    title = "{Nonlinear $\sigma$ Models and Their Gauging in and Out of Superspace}",
    reportNumber = "STOCKHOLM-ITP-85-2",
    doi = "10.1016/0550-3213(86)90175-6",
    journal = "Nucl. Phys. B",
    volume = "266",
    pages = "1--44",
    year = "1986"
}

@article{Uvarov:2017lyu,
    author = "Uvarov, D. V.",
    title = "{Massless spinning particle and null-string on AdS$_d$ : projective-space approach}",
    eprint = "1707.05761",
    archivePrefix = "arXiv",
    primaryClass = "hep-th",
    doi = "10.1088/1751-8121/aac5d3",
    journal = "J. Phys. A",
    volume = "51",
    number = "28",
    pages = "285402",
    year = "2018"
}

\end{document}